\documentstyle[aps,preprint]{revtex} 
\def\P{\Phi}
\def\p{\phi}
\def\s{\psi}
\def\m{\mu}
\def\r{\rho}
\def\c{\chi}
\def\n{\nu}
\def\a{\alpha}
\def\b{\beta}
\def\g{\gamma}

\def\l{\lambda}
\def\pa{\partial}
\def\d{\delta}
\def\e{\epsilon}
\def\i{\bibitem}
\def\be{\begin{eqnarray}}
\def\ee{\end{eqnarray}}
\def\nn{\nonumber}
\def\f{\frac}
\def\ll{\left.}
\def\rr{\right.}

\begin{document}
\draft
\title
{\bf \LARGE Supersymmetric  Gauged $ O(3) $ Sigma Model and Self-dual
Born-Infeld Theory}
\narrowtext
\author{Prasanta K. Tripathy\thanks{E-mail: prasanta@iopb.res.in}}
\address{Institute of Physics, Bhubaneswar 751005, India}
\maketitle

\begin{abstract}
We study the supersymmetric extension of the gauged $ O(3) $ sigma model
in $ 2+1 $ dimensions and find the supersymmetry algebra. We also discuss 
soliton solutions in case the Maxwell term is replaced by the Born-Infeld term.
We show that by appropriate choice of the potential, the self-dual equations
in the Born-Infeld case coincide with those of the Maxwell's case.
\end{abstract}

\section{Introduction}
The $ O(3) $ sigma model in $ 2+1 $ dimensions is well studied
\cite{Raj}' \cite{plo}. It is well known that this model is integrable 
and has Bogomol'nyi type bound on energy\cite{bogom}. All finite energy static 
solutions of the Bogomol'nyi equations are well known and can be expressed 
in terms of rational functions \cite{Raj}' \cite{plo}. 
The model is scale invariant and hence the solitons are of arbitrary 
size. Bogomol'nyi type  bound has also been obtained by 
adding a suitable potential to the model so as to break the scale 
invariance\cite{leese}. Some time ago, gauged $ O(3) $ sigma model was  
also well studied when the gauge field dynamics is governed by a Maxwell 
term\cite{schor} and Bogomol'nyi bound was obtained. It was shown that
these solitons are topologically stable even though they carry arbitrary 
value of magnetic flux. Around the same time, people also considered
gauged $ O(3) $ sigma model when the dynamics is governed by the Chern-Simons 
term alone and all static solutions were obtained for both abelian as well as 
non-abelian gauge fields \cite{Nard}'\cite{cho}. Gauged $ O(3) $ sigma models 
in the presence of both the Maxwell and the Chern-Simons
term also admit static self-dual solutions when anomalous magnetic 
moment interaction is incorporated\cite{pijush}. 

It is well known that self-duality is closely related to $ N = 2 $ 
supersymmetry\cite{lee}. Supersymmetric pure $ O(3) $ sigma model ( without 
gauge field ) was indeed constructed\cite{vecchia} long time ago. Recently
supersymmetric verson of the gauged sigma model has also been constructed 
in case the gauge field dynamics is governed by the abelian Chern-Simons term
\cite{kimm}. However, to the best of our knowledge no supersymmetric model 
has been constructed for gauged sigma model with Maxwell term. The aim of 
this paper is to carry out this task.

Recently Born-Infeld theory \cite{Born},\cite{Infeld} has received wide 
publicity, specially in the context of string theory\cite{String}. 
Different models containing Born-Infeld Lagrangian are found to possess 
domain wall, vortex and monopole solutions \cite{nakamura}. Supersymmetry and 
BPS saturated solutions in connection with D-brane dynamics \cite{Sudomwl} 
(which contains Born-Infeld action ) have also been investigated. 
Recently self-dual solutions have been obtained in abelian Higgs model in
$ 2+1 $ dimensions in case the Maxwell term is replaced by the abelian 
Born-Infeld term \cite{schapo}. In particular, it has been shown that by 
suitably modifying the Higgs potential, the self-dual equations for the 
Born-Infeld abelian Higgs model are identical to that of the corresponding 
Maxwell abelian Higgs model. It is then worthwhile enquiring if one can also 
obtain self-dual solutions in the gauged $ O(3) $ sigma model in case the 
Maxwell term is replaced by the Born-Infeld action. In this paper we show 
that the answer to the question is yes. In particular, we show that the 
self-dual equations are identical to the gauged $ O(3) $ sigma model with 
Maxwell term provided the potential is adjusted suitably.

The plan of the paper is the following. In section II we construct the 
supersymmetric verson of the gauged $ O(3) $ sigma model with Maxwell term
and find the corresponding supersymmetry algebra. In section III we consider 
the gauged $ O(3) $ sigma model with the Born-Infeld action and obtain the 
Bogomol'nyi bound. Finally we conclude the results in section IV and point
out some of the open problems. 

\section{Supersymmetric Gauged  Sigma Model }

We consider three component real superfield $\P ^a $ containing scalar
field $ \p ^a $ , Majorana spinor $ \s ^a $ and auxiliary field $ F^a $ in 
$ 2+1 $ dimensions. The real spinor gauge superfield $ W ^{\a } $ contains 
a gauge field $ A_{\m } $ and a Majorana spinor $ \l ^{\a } $ . A real scalar 
superfield $ S $ consists of a real scalar $ M $ , a Majorana fermion $ \c $ 
and a real auxiliary field $ D $ . The superfield $ \P ^a $ is constrained to
satisfy the relation $ \P ^a \P ^a = 1 $ , $ a = 1, ... , 3 $ which yields 
the following three constraints

\be
&& \p ^a\p ^a = 1 \nn \\
&& \p ^a\s ^a = 0 \nn \\
&& \p ^aF^a + \f{1}{2}\bar \s ^a\s ^a = 0 .
\ee

The action in terms of component fields can be written as 

\be
S = && \int d^3x \left\{ -\f{1}{4}F_{\m \n }F^{\m \n } 
+\f{1}{2}\pa _{\m }M\pa ^{\m }M 
+\f{1}{2}D_{\m }\p ^aD^{\m }\p ^a  \nn \rr \\
&& \ll + \f{1}{2} F^a F^a + \f{1}{2} D^2(x) 
+ D(x)(1 -  n^a\p ^a) \nn \rr \\
&& \ll + \f{i}{2} \bar \l \g _{\m }\pa ^{\m }\l 
+ \f{i}{2} \bar \c \g _{\m }\pa ^{\m }\c 
+ \f{i}{2} \bar \s ^a\g _{\m }D ^{\m }\s ^a\nn \rr \\
&& \ll - \bar \l \s ^a (\hat n \times \p )^a
- M n^a F^a - n^a \bar \s ^a \c \right\}
\ee

where we have 
\be
D_{\m }\p ^a = \pa _{\m }\p ^a + \e ^{abc} A_{\m }n^b\p ^c \nn .
\ee
The auxiliary fields $ F^a $ and $ D $ can be removed from the action 
by using their equations of motion 

\be
&& D = - (1 - n^a\p ^a) \nn \\
&& F^a = (n^a - \p ^an^b\p ^b)M - \f{1}{2}\p ^a \bar \s ^b\s ^b
\ee

Note that the constraints in Eq.(1) have been used to obtain the second 
of the above equations. Using Eq.(3) the action (2) becomes 

\be
S = && \int d^3x \left\{ -\f{1}{4}F_{\m \n }F^{\m \n } 
+\f{1}{2}\pa _{\m }M\pa ^{\m }M 
+\f{1}{2}D_{\m }\p ^aD^{\m }\p ^a  \nn \rr \\
&& \ll + \f{i}{2} \bar \l \g _{\m }\pa ^{\m }\l 
+ \f{i}{2} \bar \c \g _{\m }\pa ^{\m }\c 
+ \f{i}{2} \bar \s \g _{\m }D ^{\m }\s \nn \rr \\
&& \ll - \f {1}{2}(1 - n^a\p ^a)^2 
- \f{1}{2}M^2\left[ 1 - (n^b\p ^b)^2\right] \nn \rr \\
&&\ll - \bar \l \s ^a(\hat n \times \p )^a
- n^a\bar \s ^a\c \nn \rr \\
&& \ll + \f{1}{2}Mn^b\p ^b\bar \s ^c\s ^c 
+ \f{1}{4}(\bar \s ^b\s ^b)^2 \right\}
\ee

The action(4) is invariant under the following $ N=2 $ supersymmetry 
transformations

\be
&& \d \p ^a = \bar \e \s ^a \nn \\
&& \d A_{\m } = i\bar \e \g _{\m }\l \nn \\
&& \d M = \bar \e \c \nn \\ 
&& \d \l = \f{i}{2}\e ^{\m \n \r }F^{\m \n }\g ^{\r }\e \nn \\
&& \d \c = -i\g _{\m }\pa ^{\m}M\e + (1 - n^a\p ^a)\e \nn \\
&& \d \s ^a = -i\g _{\m }D^{\m }\p ^a\e - \left[ 
(n^a - \p ^an^b\p ^b)M - \f{1}{2}\p ^a \bar \s ^b\s ^b
\right]\e 
\ee

The corresponding supercharges are 
\be 
Q^1 = && \int d^2x \left[\g ^{\m }\g ^0\s ^aD_{\m }\p ^a
+ i\g ^0\s ^a n^aM \nn \rr \\
&& \ll - \e _{\r \m \n }\g ^{\r }\g ^0\l \pa ^{\m }A^{\n }
+ \g ^{\m }\g ^0\c \pa _{\m }M 
+ i \g ^0\c (1 - n^a\p ^a)\right]
\ee

and 

\be
Q^2 = && \int d^2x \left[(\p \times D_{\m }\p )^a \g ^{\m }\g ^0\s ^a
- i (n\times \p )^a\g ^0\s ^aM \nn \rr \\
&& \ll - \e _{\r \m \n }\g ^{\r }\g ^0\c \pa ^{\m }A^{\n }
+\g ^{\m }\g ^0\l \pa _{\m }M + i\g ^0\l (1 - n^a\p ^a)\right]
\ee

which satisfy the following anticommutation relations

\be
\f{1}{2}\left \{Q_{\a }^A , Q_{\b }^B\right \}
=  (\g _0)_{\a \b }P_0\d ^{AB} + T\e _{\a \b }\e ^{AB}
\ee

where $P_0  $, the static energy is given by 

\be
\int d^2x \left[\f{1}{4}F_{ij}F^{ij} 
+ \f{1}{2}\mid D_i\p \mid ^2 
+ \f {1}{2}(1 - n^a\p ^a)^2\right]
\ee

and the central charge 

\be 
T = \int d^2x \left[\e ^{abc}\p ^aD_1\p ^bD_2\p ^c 
+ F_{12}(1 - n^a\p ^a) \right]
\ee

Note that after calculating the superalgebra we restrict the model to its
bosonic sector by putting the fermion fields to zero.
As expected, the above expression for static energy coincides with that 
in ref.\cite{schor} and the central charge is identical to the topological 
charge derived in that paper.
       Note that the anticommutator in Eq.(8) is hermitian and hence
the trace of its square is positive semi-definite.

\be
\sum _{AB}\{Q_\a ^{(A)},Q^{\b (B)}\}\{Q^{\a (A)},Q_{\b }^{(B)}\} \ge 0
\ee

\be
\Rightarrow P_0 \ge \mid T \mid
\ee

which is the Bogomol'nyi bound on the energy.

\section{The Born-Infeld Gauged $ O(3) $ Model}

Now we construct the action for the gauged $ O(3) $ sigma model with the 
Born-Infeld term. Taking a clue from the abelian-Higgs model with the
Born-Infeld term \cite{schapo}, here we introduce a potential which 
appears as a multiplicative factor inside the square root of the 
Born-Infeld term. We start with the following action for scalar fields 
with a Born-Infeld term

\be
S = \int d^3x \left[\f{1}{2}D_{\m }\p ^aD^{\m }\p ^a + \b ^2
- \b ^2\sqrt{\left (1 + \f{1}{2\b ^2}F_{\m \n }F^{\m \n }\right )V(\p )}
\hspace{.2cm}\right]
\ee

where $ V(\p ) $ is given as 

\be
V(\p ) = 1 + \f{1}{\b ^2}\left (1 - n^a\p ^a\right )^2 
\ee

As before we restrict the field $ \p ^a $ be $ \p ^a\p ^a = 1 $.
Note that in the limit $ \b \rightarrow \infty $ the above action 
reduces to 

\be
S = \int d^3x \left[ - \f{1}{4}F_{\m \n }F^{\m \n }
+ \f{1}{2}D_{\m }\p ^aD^{\m }\p ^a 
- \f{1}{2}(1 - n^a\p ^a)^2 \right]
\ee

which is the action for the gauged $ O(3) $ sigma model with the Maxwell
term. 

The eqauations of motion are 

\be
&& D_{\m }D^{\m }\p ^a - (\p ^bD_{\m }D^{\m }\p ^b)\p ^a \nn \\
&& + (n^a - \p ^an^b\p ^b)(1 - n^c\p ^c)
\sqrt{\f{1 + \f{1}{2\b ^2}F_{\m \n }F^{\m \n }}{V(\p )}} = 0
\ee

and 

\be
\pa _{\m }\left (F^{\m \n }
\sqrt{\f{V(\p )}{1 + \f{1}{2\b ^2}F_{\m \n }F^{\m \n }}}
\hspace{.2cm}\right ) + \e ^{abc}D^{\n }\p ^an^b\p ^c = 0 .
\ee

The static energy is found to be 

\be
E = \int d^2x \left( \f{1}{2}(D_i\p ^a)^2
+ \b ^2\left [\sqrt{\left (1 + \f{F_{12}^2}{\b ^2}\right )
\left (1 + \frac{1}{\b ^2}(1 - n^a\p ^a)^2 \right )} - 1\right ]\right)
\ee

The energy can be rearranged as 

\be
E & = & \int d^2x \left(\f{1}{2}(D_1\p ^a + \e ^{abc}\p ^bD_2\p ^c)^2
+ \e ^{abc}\p ^aD_1\p ^bD_2\p ^c \nn \rr \\
&& \ll + \b ^2\left [\sqrt{\f{1}{\b ^2}
\left \{F_{12} - (1 - n^a\p ^a)\right \}^2
+ \left \{1 + \f{1}{\b ^2}F_{12}(1 - n^a\p ^a)\right \}^2
} - 1\right ]\right)
\ee

From the above expression we get the following bound on the
energy

\be
E \ge \int d^2x \left(\e ^{abc}\p ^aD_1\p ^bD_2\p ^c + F_{12}(1 - n^a\p ^a)
\right)
\ee

which is same as that for the Maxwell case.
The bound is saturated when the fields satisfy the Bogomoln'yi 
equations 

\be
&& D_1{\p ^a} + \e ^{abc}\p ^b D_2\p ^c = 0 \nn \\
&& F_{12} - (1 - n^a\p ^a) = 0
\ee

\noindent
which are identical to the Bogomol'nyi equations for the Maxwell's 
case \cite{schor}. Note that these are also 
consistent with the second order field equations Eqs. (16-17) .
Hence all the properties like flux and charge for this model will
be same as those for the Maxwell's case. 

It is well known that results similar to $ O(3) $ sigma model also 
holds in $ CP^1 $ model. As a confirmation of the fact, we now obtain
a similar bound on energy in the case of gauged $ CP^1 $ model with 
Born-Infeld action.
Let $ \xi $   be the $ CP^1 $ field such that $ \xi ^{\dagger }\xi = 1 $ .
Then consider the following action

\be
S = \int d^3x \left[ \mid \nabla _{\mu }\xi\mid ^2 + \b ^2
- \b ^2\sqrt{\left (1 + \f{1}{2\b ^2}F_{\m \n }F^{\m \n }\right )V(\xi )}
\hspace{.2cm}\right]
\ee

where 

\be 
\nabla _{\mu }\xi = D_{\mu }\xi - \left(\xi ^{\dagger }D_{\mu }\xi \right)\xi
\ee

\be
V(\xi ) =  1 + \f{1}{\b ^2}\left (1 - \xi ^{\dagger }\sigma _3\xi \right )^2
\ee

and 

\be
(D_{\mu }\xi )^T 
= \left([\pa _{\mu } - iA_{\mu }]\xi _1 ,\pa _{\mu }\xi _2\right)
\ee

The static energy is 
\be
E = \int d^2x \left( \mid\nabla _i\xi\mid ^2
+ \b ^2\left [\sqrt{\left (1 + \f{F_{12}^2}{\b ^2}\right )
\left (1 + \frac{1}{\b ^2}(1 - \xi ^{\dagger }\sigma _3\xi)^2 \right )} 
- 1\right ]\right)
\ee

Using the identity 
\be 
\mid\nabla _i\xi\mid ^2 = \mid (\nabla _1 + i\nabla _2)\xi\mid ^2
+ i\e _{jk}(\nabla _k\xi )^\dagger (\nabla _j\xi )
\ee

the energy can be rewritten as ,

\be
E & = & \int d^2x \left( \mid (\nabla _1 + i\nabla _2)\xi\mid ^2
+ i\e _{jk}(\nabla _k\xi )^\dagger (\nabla _j\xi ) \nn \rr \\
&& \ll + \b ^2\left [\sqrt{\f{1}{\b ^2}
\left \{F_{12} - (1 - \xi ^{\dagger }\sigma _3\xi)\right \}^2
+ \left \{1 + \f{1}{\b ^2}F_{12}(1 - \xi ^{\dagger }\sigma _3\xi)\right \}^2
} - 1\right ]\right)
\ee

and hence we have the inequality

\be
E \ge \int d^2x \left(i\e _{jk}(\nabla _k\xi )^\dagger (\nabla _j\xi )
+ F_{12}(1 - \xi ^{\dagger }\sigma ^3\xi )\right)
\ee

The above bound is saturated if the following Bogomol'nyi equations hold.

\be
&&\nabla _1\xi + i\nabla _2\xi = 0 \\
&& F_{12} - (1 - \xi ^{\dagger }\sigma _3\xi ) = 0
\ee

which are the self-dual equations for the gauged $ CP^1 $ model with
Maxwell's term.

\section{Conclusion}
In this paper we have constructed a supersymmetric extension of the gauged 
$ O(3) $ sigma model with the Maxwell term. Replacing the 
Maxwell term by the Born-Infeld term in the above model
we have obtained the self-dual equations and have shown that by changing the
potential appropriately, these self-dual equations are same as in the 
Maxwell's case. This work raises several problems which deserves further study.
Some of these are \\
(i) can one find supersymmetric extension of the gauged $ O(3) $ sigma
model with both the Maxwell and the Chern-Simons terms? \\
(ii) In case of abelian Higgs model the maximal supersymmetry \cite{hckao}
is $ N = 3 $. Thus it is natural to enquire about the maximal supersymmetry 
in the case of sigma model with Maxwell term. \\
(iii) Can one obtain self-dual solutions in the gauged sigma model with both 
Chern-Simons and Born-Infeld action? \\
(iv) Further can one obtain the supersymmetric version of the above? \\
We hope to report on some of these issues in the near future.

\section{Acknowledgments}

I am indebt to Avinash Khare for helpful discussins as well as for careful
reading of the manuscript.

\end{document}